\documentclass[useAMS,usenatbib,letterpaper]{mn2e}
\usepackage{amsmath,amssymb}
\usepackage{graphicx}
\usepackage{natbib}
\usepackage[usenames]{color}
\def\Omm{{\Omega_m}}
\def\Ommz{{\Omega_m^{\,z}}}

\def\Omk{{\Omega_k}}
\def\Oml{{\Omega_{\Lambda}}}

\def\aap{A\&A}
\def\apj{ApJ}

\def\apjl{ApJ}
\def\mnras{MNRAS}
\def\araa{ARA\&A}
\def\aj{AJ}

\def\prd{Phys. Rev. D}
\def\nat{Nat}

\def\apjs{ApJS}
\newcommand{\beq}{
\begin{equation}
}
\newcommand{\eeq}{
\end{equation}
}
\newcommand{\kms}{\,{\rm km\,s^{-1}}}
\newcommand{\msun}{\,{\rm M_\odot}}

\def\spose#1{\hbox to 0pt{#1\hss}}
\newcommand{\lta}{\mathrel{\spose{\lower 3pt\hbox{$\mathchar"218$}}
    \raise 2.0pt\hbox{$\mathchar"13C$}}}
\newcommand{\gta}{\mathrel{\spose{\lower 3pt\hbox{$\mathchar"218$}}
    \raise 2.0pt\hbox{$\mathchar"13E$}}}
\def\simlt{\mathrel{\rlap{\lower 3pt\hbox{$\sim$}}\raise 2.0pt\hbox{$<$}}}
\def\simgt{\mathrel{\rlap{\lower 3pt\hbox{$\sim$}} \raise 2.0pt\hbox{$>$}}}

\DeclareMathAlphabet{\mathvecbf}{OT1}{ptm}{bx}{it}
\newcommand{\vecb}[1]{\ensuremath{\mathvecbf {#1}}}
\newcommand{\units}[1]  {\ensuremath{\mathrm{~{#1}}}}

\title[Rocket science] {
Gravitational recoil: effects on massive black hole occupation fraction over cosmic time
}

\author[Volonteri et al.]  {Marta Volonteri$^{1}$, Kayhan
G\"{u}ltekin$^{1}$, and Massimo Dotti$^{1,2}$\\ 
$^1$Department of Astronomy, University of Michigan, Ann Arbor, MI, 48109, USA\\ 
$^2$Max-Planck-Institut f\"{u}r Astrophysik, Garching bei M\"{u}nchen, 85740, Germany}
\pagerange{\pageref{firstpage}--\pageref{lastpage}}
\definecolor{KayhanCiteColor}{rgb}{0,0.08,0.35}
\definecolor{KayhanURLColor}{rgb}{0,0.08,0.35}
\definecolor{KayhanLinkColor}{rgb}{0,0.08,0.35}
\definecolor{KayhanPageColor}{rgb}{0,0.08,0.35}

\begin{document}

\label{firstpage}

\maketitle

\begin{abstract}
We assess the influence of massive black hole (MBH) ejections from
galaxy centres due to gravitational radiation recoil, along the cosmic
merger history of the MBH population. We discuss the `danger' of
recoil for MBHs as a function of different MBH spin-orbit
configurations and of the host halo cosmic bias, and on how that
reflects on the occupation fraction of MBHs. We assess ejection
probabilities for mergers occurring in a gas-poor environment, in
which the MBH binary coalescence is driven by stellar dynamical
processes and the spin-orbit configuration is expected to be
isotropically distributed. We contrast this case with the `aligned'
case. The latter is the more realistic situation for gas-rich, i.e.,
`wet,' mergers, which are expected for high-redshift galaxies.
We find that if {\it all} haloes at $z>5$--$7$ host a MBH, the probability
of the Milky Way (or similar size galaxy) to host a MBH today is less
than 50\%, unless MBHs form continuously in galaxies.  The occupation
fraction of MBHs, intimately related to halo bias and MBH formation
efficiency, plays a crucial role in increasing the retention
fraction. Small haloes, with shallow potential wells and low escape
velocities, have a high ejection probability, {\it but} the MBH merger
rate is very low along their galaxy formation merger hierarchy: MBH
formation processes are likely inefficient in such shallow potential
wells.  Recoils can decrease the overall frequency of MBHs in small
galaxies to $\sim60$\%, while they have little effect on the frequency
of MBHs in large galaxies (at most a 20\% effect).
\end{abstract}
\begin{keywords} black hole physics --
galaxies: formation -- 
cosmology: theory 
\end{keywords}

\section{Introduction}
It is well established that galaxies contain massive black holes
(MBHs) in their nuclei \citep[e.g.,][]{kr95, richstoneetal98}.
Observations of the host elliptical galaxy or spiral bulge show a
relation between the mass of the MBHs and the galactic spheroid
luminosity and the stellar velocity dispersion
\citep{magorrianetal98,fm00,gebhardtetal00a}.  Such relationships, as
well as their small instrinsic scatter
\citep{tremaineetal02,gultekinetal09b}, hint at a connection between
the formation of the MBH and the formation of the spheroid.

In the context of the favored cold dark matter cosmology, hierarchical
assembly of galaxies and protogalaxies with MBHs at their centers
naturally leads to the formation of MBH binaries \citep{bbr80}.  If
the MBH binary can achieve a small enough separation through stellar
dynamical hardening or through viscous gas dynamical drag,
gravitational radiation from the binary can become significant enough
to cause the system to merge.  Gravitational waves carry away energy
from the binary system, driving the system to merger. If the black
holes have unequal masses or misaligned spins, anisotropic
gravitational radiation will impart a net momentum flux on the binary,
causing the center of mass to recoil.

Until recently, the magnitude of the recoil velocity was uncertain.
For non-spinning black holes the maximum recoil has now been
calculated to be $v_{\rm recoil,max} \approx 200\kms$
\citep{bakeretal06}, and a similar range is expected for black holes
with low spins, or with spins (anti-)aligned with the binary orbital
angular momentum.  Several studies have also found consistent results
for the maximum recoil of spinning black holes $v_{\rm recoil,max} =
1000$--$4000\kms$ \citep{bakeretal07,gonzalezetal07,herrmannetal07,
koppitzetal07,campanellietal07,sb07}.  Such velocities are significant
because they will eject the binary from the galaxy since most galaxies
have escape speeds $ < 1000\kms$.

\cite{vhg2008} studied the effect of recoil on the MBH occupation
fraction in nearby galaxies.  They assumed that the relative
orientation between the orbital angular momentum of the binary and the
spins of the two MBHs were isotropically distributed.  This
configuration can result in high recoil velocities, up to thousands
$\kms$.  However, \cite{brm07} proposed that MBHs orbiting in gaseous
circumnuclear discs, such as those expected in advanced stages of gas
rich galaxy mergers \citep{mayer2007}, align their spins with the
orbital angular momentum of the binary. This configuration leads to
small recoils for the MBH remnant.  The accreting gas exerts
gravito-magnetic torques that suffice to align the spins of both the
MBHs with the angular momentum of the large-scale gas flow.
\cite{dotti2009b} have quantified the efficiency of this alignment
process through the analysis of high resolution {\it N}-body Smoothed
Particle Hydrodynamics (SPH) simulations.  They apply the algorithm
presented in \cite{perego2009} to evolve masses, magnitudes and
orientation of the spins of two MBHs.

We extend the investigation of \cite{vhg2008}
considering different degrees of alignment between the spins and the
angular momentum of the MBHs.  We will either use isotropic
spins and angular momentum configurations or the `quasi-aligned'
distributions of spin orientations from \cite{dotti2009b}. Those two
cases bracket all the possible values of recoil velocities.
We apply these distributions to Monte--Carlo realizations of galaxy
merger trees, and we study the effect of gravitational recoil on the
occupation fraction of MBHs in galaxies at different redshifts.

\section{Recoil Velocities}

\subsection{Fitting formulae}
The recoil velocity may be broken down into components arising solely
from mass asymmetry, $v_m$, which is perpendicular to the orbital
angular momentum vector $\vecb{L}_{\rm pair}$, and components arising
from spin asymmetry, $v_\perp$ and $v_\parallel$, which are
perpendicular and parallel to $\vecb{L}_{\rm pair}$, respectively:
\beq
{v}_{\rm recoil} = \sqrt{v_m^2 + v_{\perp}^2+2 v_m v_{\perp} \cos(\xi)+ 
    v_{\parallel}^2}, 
\label{eq:v_total}
\eeq
where $\xi$ is the angle between $v_m$ and $v_\perp$ in the orbital
plane.  We assumed $\xi=145^{\circ}$, as suggested by \cite{lz2009}.
\cite{campanellietal07} and \citet[][fit CL]{lz2009}
propose the following fitting formulas for the recoil components:
\begin{eqnarray}
v_m &=& A \eta^2 \sqrt{1 - 4 \eta}\, (1 + B \eta), \label{eq:v_mass}\\ 
v_{\perp} &=& H \eta^2(1+q)^{-1}\left( a_1^{\parallel} - q a_2^{\parallel}
    \right), \label{eq:v_perp}\\ 
v_{\parallel} &=& K \eta^2(1+q)^{-1}\,\cos(\Theta-\Theta_0)\left|\vecb{a}_1^{\perp} - 
    q \vecb{a}_2^{\perp}\right|, \label{eq:v_parallel}
\end{eqnarray}
where $q = M_2 / M_1 \le 1$ is the mass ratio of the black holes and
$\eta\equiv q/(1+q)^2$ is the symmetric mass ratio.  The components of
the spins of the two MBHs are broken into projections parallel and
perpendicular to $\vecb{L_{\rm pair}}$, $\vecb{a}^{\perp} =
\vecb{a}\sin(\theta)$ and $a^{\parallel} = \vecb{a}\cos(\theta)$.  In
the isotropic case, $\cos(\theta_1)$ and $\cos(\theta_2)$ are
distributed uniformly between $-1$ and $1$. We compare the results
from the isotropic case with those obtained assuming quasi--aligned
spin-orbit configurations. In this case we assumed $\theta_1$ and
$\theta_2$ were distributed according to \citet[][see
Section~\ref{sec:angles}]{dotti2009b}.  $\Theta$ is the angle between
$(\vecb{a}_2^{\perp} - q \vecb{a}_1^{\perp})$ and the separation
vector at coalescence, and $\Theta_0$ depends on the initial
separation between the holes.  We assume a uniform distribution of
$\Theta-\Theta_0$ between 0 and $2\pi$.  Note that for
$q\vecb{a}_{1\perp} = \vecb{a}_{2\perp}$ (including $a_{1\perp} =
a_{2\perp} = 0$), $v_\|$ vanishes and there is no recoil out of the
orbital plane.  In the absence of spins, the recoil is $v_m$, which is
maximized for for $q = (3 - \sqrt{5})/2 \approx 0.38$.  The best fit
parameters found by \citet[which we refer to as fit CL]{lz2009}  are $A
= 1.2 \times 10^4 \kms$, $B = -0.93$, $H = 6900\kms$, and $K = 6.0
\times 10^4 \kms$.

\citet[referred to as fit B]{bakeretal2008} found an alternative form
for $v_{\parallel}$ that scales as $\eta^3$: \beq v_{\parallel} = K
\eta^3(1+q)^{-1}\,\left(a_1^{\perp} \cos(\Phi_1)- q a_2^{\perp}
\cos(\Phi_2)\ \right), \eeq
where $\Phi_1$ and $\Phi_2$ are the differences between two angles
that depend on the system of reference. We assume $\Phi_1=\Phi_2$
uniformly distributed between 0 and $2\pi$
\citep[see][]{dotti2009b}. The best-fit parameters from
\citet{bakeretal2008} are $A = 1.35 \times 10^4 \kms$, $B = -1.48$, $H
= 7540 \kms$, and $K = 2.4 \times 10^5 \kms$.

\citet[fit H in the following]{herrmannetal07} parameterize the
recoil velocity in terms of $\theta_{\rm H}$, the angle between
$\vecb{L}_{\rm pair}$ and
\begin{equation}
{\mathbf{\Sigma}}=\left(M_1 + M_2\right)\left(\frac{\vecb{J}_1}{M_1} - \frac{\vecb{J}_2}{M_2} \right),
\end{equation}
where the $\vecb{J}$ are the spin angular momenta of the holes.
Taking $\vecb{L}_{\rm pair}$ to be in the $z$ direction, they find the
Cartesian components of the recoil velocity to be:
\begin{eqnarray}
\nonumber V_x &=& C_0 H_x \cos(\theta_{\rm H}),   \\
\nonumber V_y &=& C_0 H_y \cos(\theta_{\rm H}),   \\
        V_z &=& C_0 K_z \sin(\theta_{\rm H}),   
\end{eqnarray}
where $C_0=\Sigma q^2 (M_1 + M_2)^{-2} (1 + q)^{-4}$ with the best-fit
parameters $H_x = 2.1\times 10^3$, $H_y = 7.3\times 10^3 $, and $K_z =
2.1\times 10^4$.

\begin{figure}
\includegraphics[width=0.48\textwidth]{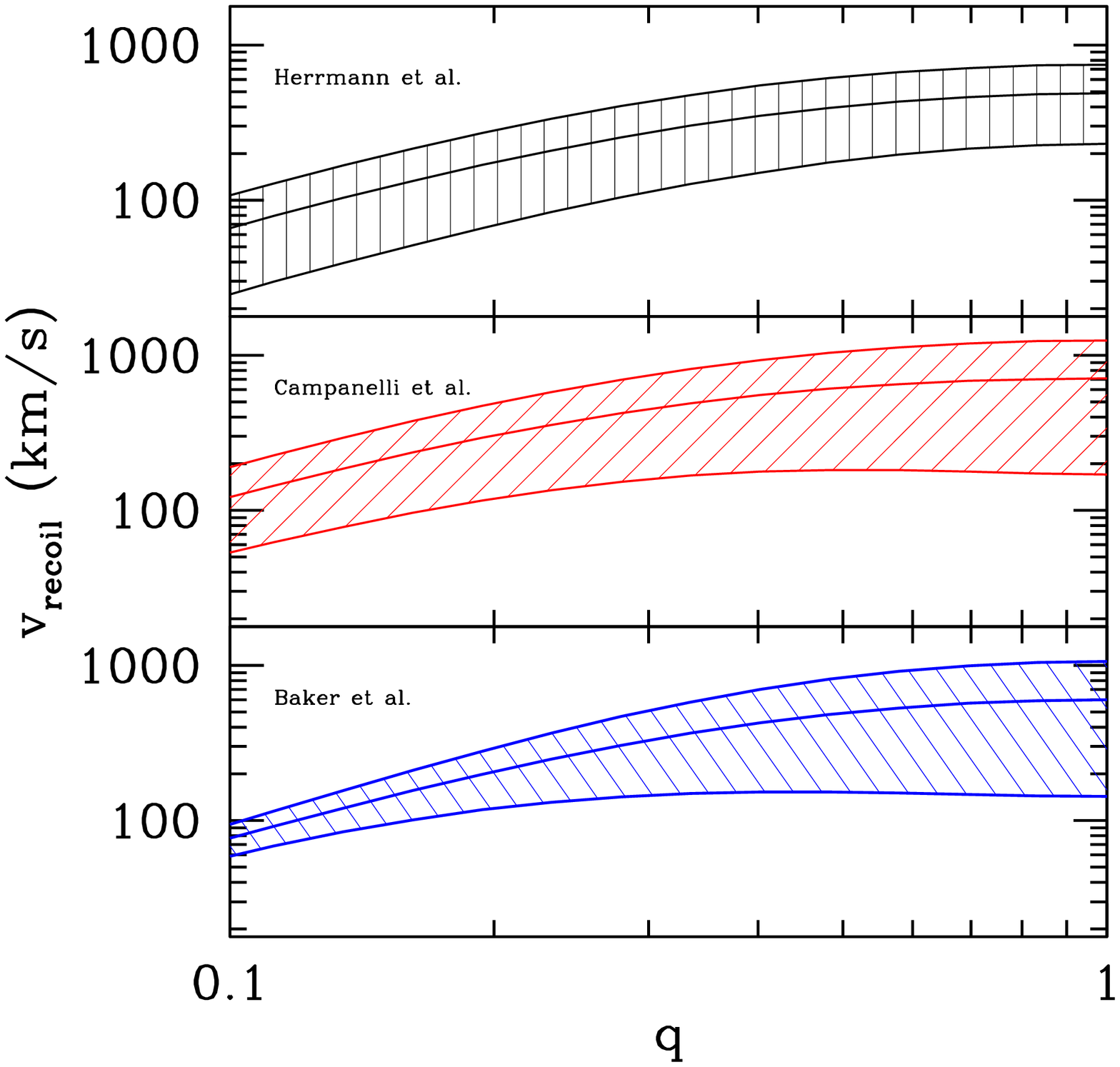}
\includegraphics[width=0.48\textwidth]{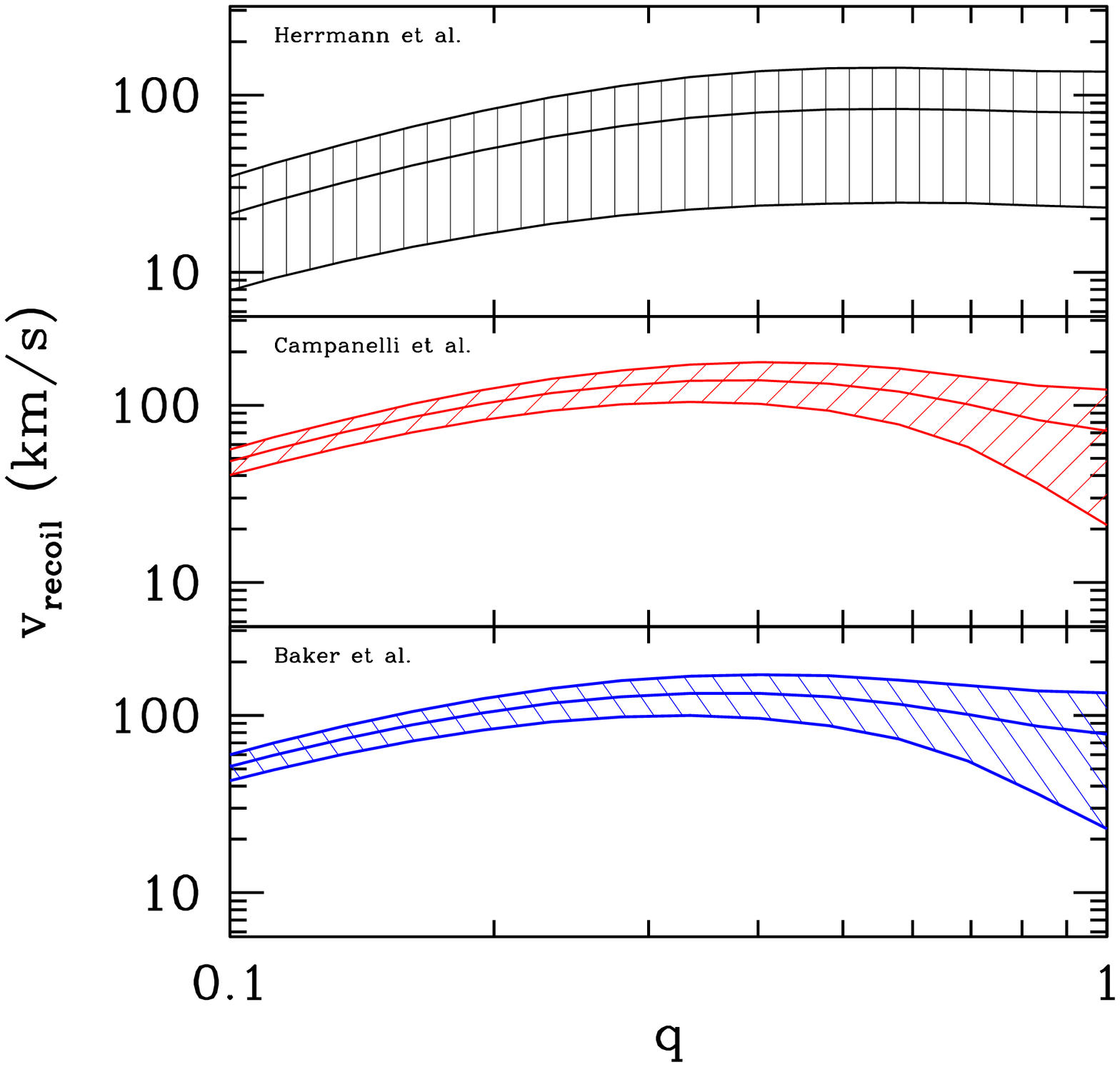}
\caption{Recoil velocity of the MBH remnant as a function of $q$, for
MBH spins isotropically distributed (top) or for aligned
configurations (bottom).  The upper, middle, and lower panel refers to
fit H, fit CL, and fit B, respectively. The thick lines show the
average values of $v_{\rm recoil}$ ($\overline{v}_{\rm recoil}$), and
$\overline{v}_{\rm recoil} \pm \sigma$.}
\label{fig:comparison}
\end{figure}

In Figure~\ref{fig:comparison} we plot recoil velocities from fits H,
CL, and B as a function of $q$ for a sample of isotropically
distributed MBH spins and for aligned configurations.  This figure
demonstrates the agreement of the three formulae for different
spin-orbit configurations.  For each value of $q$ between 0.1 and 1,
we produce 500\,000 realizations of the two MBH spins with magnitudes
uniformly distributed between 0 and 1. For the isotropic case we also
assign isotropically distributed spin directions.  We do not show
results for $q<0.1$ because the recoil velocity drops below the typical escape
velocity from galaxies.

Fit B decreases faster as $q$ decreases, with respect to fits CL and
H.  This is because the component of the recoil velocity parallel to
$\vecb{L}_{\rm pair}$ scales as $\eta^3$ in fit B, and as $\eta^2$ in
fits CL and H.  The average values of $v_{\rm recoil}$ obtained with
the different prescriptions for isotropically distributed spins are
consistent within $1\sigma$ (i.e. within a factor of $\sim 2$).
Assuming perfect alignment between the two spins and $\vecb{L}_{\rm
pair}$, the results from fits CL, H, and B are consistent within a
factor of 2.5 for any $q$.  For the aligned case with $q \approx 0.1$,
the difference between fits CL and B disappears because, for perfect
alignment, $a=a^{\parallel}$ and $v_{\parallel}=0$, resulting in a
perfect agreement between the two fits.

\subsection{Quasi-aligned configurations}\label{sec:angles}

As mentioned in the Introduction, gravo--magnetic torques exerted by
accreting flows onto the MBHs tend to align $\vecb{a}_1$ and
$\vecb{a}_2$ to $\vecb{L}_{\rm pair}$. The distributions of relative
angles between the two MBH spins, and between each spin and
$\vecb{L}_{\rm pair}$ after the formation of a MBH binary have been
computed in \citet[][see Figure~2 therein]{dotti2009b}.  Their results
stem from the analysis of high resolution {\it N}--body/SPH
simulations. In the suite of runs discussed in that paper, the Authors
varied the thermodynamical properties of the gas. In particular they
assumed two different polytropic equations of state, with polytropic
index $\gamma=5/3$ (`hot' runs) and $7/5$ (`cold' runs),
respectively.  The hot case corresponds to an adiabatic monoatomic
gas, as if radiative cooling were completely suppressed during the
merger. This case mimics gas radiatively heated by an AGN
\citep{mayer2007}. The cold case, instead, has been shown to provide a
good approximation to a gas of solar metallicity heated by a starburst
\citep[e.g.,][]{ss2000}.

In each run, \citet{dotti2009b} found a significant degree of
alignment already present before the formation of a binary.  However,
their runs always start with an equal-mass MBH pair. Because the
dynamical evolution of the two MBHs is faster than the Salpeter time,
the masses of the two MBHs do not change significantly, and at the end
of the runs a nearly equal-mass binary forms. Here we discuss the
dependence of spins/orbit configurations discussed by
\citet{dotti2009b} on the masses of the MBHs, by comparing the
dynamical timescale ($\tau_{\rm dyn}$) of the binary formation with
the alignment timescale ($\tau_{\rm align}$).  Before we consider the
ratio of these two timescales, we look at each in slightly more detail
in the context of our simulations, but for a full discussion see
\citet{dotti2009b}.

Dynamical friction is the process that drives the dynamical evolution
of the pair. Since the longer timescale is what determines the time
until coalescence, $\tau_{\rm dyn}$ is the dynamical friction
time-scale of the less massive MBH, which depends on a number of
factors:
\beq
\tau_{\rm dyn} \propto K\, M_{\rm BH}^{-1}, 
\label{eq:tdyn}
\eeq
where $K$ is a function of the properties of the circumnuclear disc.
In \citet{dotti2009b} runs, $M_\mathrm{BH} = 4 \times
 10^6\;\msun$, and $\tau_{\rm dyn}= 4.5 - 7.5$ Myr, depending on the
 initial orbital parameters of the BHs and on the effective equation
 of state used to evolve the gas thermodynamics.
Here we are considering the circumnuclear disc parameters in
\citet{dotti2009b} as fiducial.  Considering different values for
these is out of the scope of this paper, but it is likely to change
only the details and not our qualitative results.

The alignment timescale also depends on the BH spin, $a$, and may be
expressed in terms of the Eddington fraction (see equation~43 in
Perego et al. 2009):
\beq
\tau_{\rm align} \propto a^{5/7} M_{\rm BH}^{-2/35}
f_{\rm Edd}^{-32/35}.
\label{eq:talign}
\eeq
For $M_\mathrm{BH} = 4 \times 10^6\;\msun$ in the simple case of a
fixed orbital plane of the outer disc, i.e., coherent accretion, the
value of $\tau_{\mathrm{align}} \approx 10^5\;\mathrm{yr}$.  For the
simulations in \citet{dotti2009b}, however, the orbital plane of the
accreted particles is never constant, and thus the alignment timescale
will be longer: $\tau_\mathrm{align} \approx 1$--$4\;\mathrm{Myr}$,
depending on the initial orbital parameters of the BHs and the
effective equation of state used.
%

Since both black hole spins must achieve alignment with the disc
angular momentum for the spins to be aligned with each other, the
alignment timescale we are interested in is the longer of the two.
For most cases, this will be the timescale for the less massive
binary.  Before the formation of a binary, the more massive of the two
MBHs always accretes more mass, and has the shortest alignment
timescale \citep{dotti2009a}. Thus, we may use $M_{\rm BH} = M_2$ in the
expression for $\tau_{\rm align}$.  

The situation after the formation of a hard binary, however, is
slightly different.  In this case, the torques exerted by the binary
onto the gas carve out a low-density region in the middle of the disc,
and the secondary MBH, being closer to the outer and denser gas
region, can grow faster than the primary
\citep{hayasaki2007,cuadra2009}, further aligning its spin 
with $\vecb{L}_{\rm pair}$.  For our purposes, we conservatively
assume that after forming a binary,
 the spin-orbit configuration at the time of binary formation 
does not significantly evolve.  

The ratio of these two quantities is obtained by dividing
Eq~\ref{eq:talign} by Eq~\ref{eq:tdyn} and shows the dependence on the
accretion behaviour of the two MBHs:
\beq 
\tau_{\rm align}/\tau_{\rm dyn} \sim a^{5/7} M_2^{-2/35}
f_\mathrm{Edd}^{-32/35}.
\eeq
For a fixed value of $a$ and rewriting in terms of $f_\mathrm{Edd}
\sim \dot{M} M_2^{-1}$, this expression becomes
\beq 
\tau_{\rm align}/\tau_{\rm dyn} \sim \dot{M}_2^{-32/35} M_2^{13/7}.
\eeq
If they are accreting at the Bondi rate, $\dot{M}_2\propto M_2^2$, and
consequently $\tau_{\rm align}/\tau_{\rm dyn} \propto M_2^{1/35}$.  In
this case the ratio is almost independent of the mass of the
secondary, and the assumption that the two MBHs are almost aligned is
always correct.  

If, instead, the secondary is accreting at the Eddington limit,
$\dot{M_2}\propto M_2$.  From \citet{dotti2009a}, we know that for
$M_2 \approx 4\times 10^6\;\msun$, $\tau_{\rm align}/\tau_{\rm dyn}
\approx 0.3$.\footnote{We took the cold, prograde case as our
fiducial case, but the results are not very different for the other
cases.}  Using this as our normalization, we get $\tau_{\rm
align}/\tau_{\rm dyn} \approx 0.1 \, M_{2,6}^{33/35}$, where $M_{2,6}$
is the mass of the secondary in units of $10^6 \msun$.  This scaling
implies that a binary can form before the two MBHs align their spins
to $\vecb{L}_{\rm pair}$ only for very massive secondaries ($M_2
\simgt 2\times 10^7 \msun$) and rapidly accreting pairs.

To estimate the relevance of such mergers, we consider the prevalence
of the \emph{secondary} MBH having mass $M_2 \simgt 2\times 10^7
\msun$.  Assuming the scaling of MBH masses with the velocity
dispersion of the host determined for nearby galaxies
\citep{gultekinetal09b}, the host of a MBH with mass $M_2 \simgt 2
\times 10^7 \msun$ has $\sigma \simgt 130 \kms$, and an escape
velocity $\simgt 1000 \kms$.  The recoil velocity has a significant
probability of being $>1000\kms$ only for nearly equal-mass mergers
($q>0.3$).  Such mergers are extremely rare for the heaviest MBHs
\citep[see section 4;][]{volonteri07,GW3}.  Furthermore, as mentioned
above, further accretion can increase the degree of alignment of the
binary system.  As a consequence, in this study we may neglect the
dependence of the spin-orbit configurations on the MBH masses and
adopt the same distributions of relative angles between the two MBH
spins for all MBH mergers in our sample.  Note, also, that the
isotropic distribution case we test below provides an upper limit to
the ejection probability at all masses.

\section{Ejections and halo bias}
In models of structure formation based on gravitational instabilities
in Gaussian primordial fluctuations, the number density and bias
properties of a halo can be expressed as a function of its rms
fluctuation of the linear density field at redshift $z$,
$\sigma(M_h,z)$, and on the threshold density for collapse of a
homogeneous spherical perturbation at redshift z, $\delta_c(z)$.

In the \citet{ps74} formalism the comoving number density of haloes of
mass between $M$ and $M+dM$ can be expressed as:
\beq
\frac{dn}{dM}=\sqrt{\frac{2}{\pi}}\, \frac{\rho_m}{M}\, \frac{-d(\ln \sigma)}{dM}\,\nu_c\, e^{-\nu_c^2/2}\ , 
\eeq 
where $\nu_c=\delta_{\rm c}(z)/\sigma(M,z)$ is the number of standard
deviations which the critical collapse overdensity represents on mass
scale $M$.  At any redshift we can identify the characteristic mass
(i.e., $\nu_c=1$), and its multiples. The higher $\nu_c$, the more
massive and rarer the halo, and the higher its bias and clustering
strength. Although this formalism is derived from a Press-Schechter
analysis, it agrees fairly well with the results of {\it N}-body
simulations.

\cite{mw01} suggested that clustering analysis of quasar
samples can be deconvolved to yield the typical mass of haloes hosting
quasars, and their typical $\nu_c$. \citet{mw01} and subsequent
investigations \citep{shenetal07,myersetal07,pmn04} have found that
high redshift quasars are highly biased objects with respect to the
underlying matter, and that their $\nu_c$ increases with
$z$. \cite{shenetal07} and \cite{myersetal07} find that the bias increases from
$\nu_c\simeq 3$ at $z=2$, to $\nu_c\simeq 3.5$ at $z=3$, and
$\nu_c\simeq 5.5$ at $z=5$.

We evaluate the ejection probabilities, using the techniques described
in section 2 with fit CL, for haloes representing
peaks of the density fluctuations $\nu_c$=1, 2, 3, 4, 5, 6 as a
function of redshift in a concordance $\Lambda$CDM cosmology
\citep{spergeletal07short}. For every halo mass we estimate the
ejection probability by comparing the recoil velocity ($v_{\rm
 recoil}$) to the escape velocity from the dark matter halo potential
well (truncated at the virial radius). We model the halo potential
with a \cite{nfw97} density profile, where the halo properties evolve
as suggested by \cite{bullocketal01}.

The ejection probabilities thus defined are a lower limit to the
probability that the recoil voids a galaxy of its central MBH, as if
the recoil velocity is lower than the escape velocity but higher than
the velocity dispersion of a halo, the timescale for the recoiled MBH to
return to the center under the effect of dynamical friction is likely
longer than the Hubble time \citep[and references therein]{mq04,Gualandris2008,vm2008,Devecchi2009,Guedes2009}.

\begin{figure}
\includegraphics[width=1.0\columnwidth]{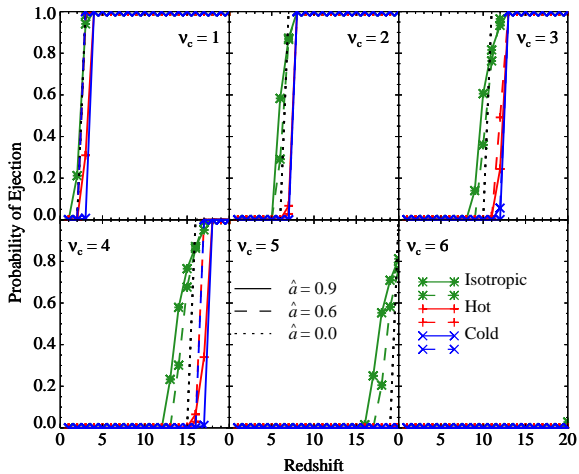}
\caption{Ejection probability as a function of redshift for different
$\nu_c=\delta_{\rm c}(z)/\sigma(M,z)$ peaks of the density
fluctuations field.  Line styles indicate the assumed spin of the
merging black holes: {\it Solid lines:} $\hat a=0.9$; {\it Dashed
lines:} $\hat a=0.6$; {\it Dotted lines:} $\hat a=0$.  Line colors
indicate assumed distribution of spin orientations: {\it Green:}
isotropically distributed; {\it Red:} orientation distribution from
hot simulations by Dotti et al.\ (2009); {\it Blue:} orientation
distribution from cold simulations by Dotti et al.\ (2009).  All
probabilities assume $q = 0.1$.}
\label{fig:ej_peaks}
\end{figure}

With this caveat in mind we can estimate lower limits to the probability that
MBH binaries are ejected or displaced due to the gravitational recoil in
$\nu_c$--peaks haloes (Figure \ref{fig:ej_peaks}). 

The redshift at which the probability drops to 50\% increases with
increasing $\nu_c$ for all binary mass ratios and spins.  The
probability of ejection for MBH binaries in haloes which host very high
redshift quasars ($\nu_c\simeq 5.5$ at $z=5$) drops to 0 at $z\gta 19$
if MBHs are non-spinning, while for spinning MBHs the ejection
probability is significant down to $z\gta 13$  in the absence of
 any alignment process. Lower $\nu_c$ peaks remain dangerous for
merging binaries for longer. A $\nu_c\simeq 3$ peak has an ejection
probability of 100\% until $z=13$. If {\it all} $\nu_c\simeq 3$ haloes
host a MBH, and {\it all} experience a major merger before $z=13$,
then no MBHs are left for further evolution. In this scenario, MBHs
must form at $z<11$, in order to evolve all the way to $z=3$ and host
the observable quasars.

However, the $\nu_c$-peak formalism does not allow a clear evaluation
of the merger history of haloes. A halo which represents a $\nu_c$-peak
at a given redshift $z$, would be incorporated into a halo
representing a lower peak at $z-\Delta z$. A MBH in a given halo would
then belong to different $\nu_c$-peaks in its lifetime. In order to
evaluate the number of MBH mergers that a halo experienced it is
necessary to use techniques that can trace the whole merger history of
a given halo as a function of mass and time.

\section{Individual halo histories}
We now turn to evaluate the merger histories of different galaxy
haloes.  We adopt a statistical approach, based on merger
histories extracted from full $\Lambda$CDM merger trees.  This work
has a more general approach than \citet{vhg2008}, and addresses in
an astrophysical context the analysis by \citet{schnittman07}.  While
\citet{vhg2008} looked at the role of ejections in the merger histories
for galaxies that were in clusters but not in the `main trunk,' here
we examine merger histories with a focus on the main halo of the
merger tree.  We evaluate the average ejection probability as a
function of cosmic epoch, and consider how it depends on the
environmental bias.
We consider here the halo merger histories leading to the formation of
haloes with masses $M_0=4\times10^{13}\msun$,
$M_0=2\times10^{12}\msun$, $M_0=2\times10^{11}\msun$ at $z=0$.  We
average our results over 20 realizations of the same mass, to account
for cosmic variance.   Our technique and cosmological framework is similar 
to the one described in \cite{VHM}. We track the dynamical evolution of  
MBHs ab-initio and follow their assembly down to $z = 0$. 

Several theoretical arguments indicate that  MBH formation occurs at very 
high redshift, and probably in biased haloes \citep[e.g.,][]{mr01,vr06}. 
Additional parameters, such as the angular momentum of the gas, its ability to cool and 
its metal enrichment, are likely to set the exact efficiency of  MBH formation and the 
redshift range when the mechanism operates \citep[for a thorough discussion see][]{VLN2008}. 
We consider the effect of varying the initial conditions, by assuming frequent or rare MBH
formation process. As a reference model, we assume that MBH formation
is effective in all haloes with $\nu_c>3$ at $z>20$ (corresponding to
masses $>10^6 \msun$). This model is based on a scenario where MBHs
form as remnants of the first generation of metal--free stars
(Population III).  The main features of the hierarchical assembly of
MBHs left over by the first stars in a $\Lambda$CDM cosmology have
been discussed by \cite{VHM}, \cite{vr06} and \cite{VN09}.  A
model similar to the one presented in this paper has been shown to
reproduce observational constraints on MBH evolution \citep[luminosity
function of quasars and Soltan's argument, $M_{\rm BH}$--$\sigma$
relationship at $z=0$, mass density in MBHs at $z=0$;][]{VLN2008}.  We
also analysed a very different case, where whenever a halo grows above
a given mass threshold ($>10^{10} \msun$) and it does not already
contain a MBH in its centre, then it forms one (in case the previous
MBH had been displaced or ejected, a new MBH materializes), regardless
of redshift.  This model is not based on a specific physical model,
but it is meant to be used for comparison with existing numerical
simulations.  MBH formation mechanisms define when and how often haloes are populated 
with MBHs. They provide the initial occupation fraction.  

Additionally we have to follow the dynamical evolution 
of haloes and embedded  MBHs all the way to $z=0$ in order to determine the 
effect of recoils on the occupation fraction of  MBHs. Since the magnitude of the recoil 
depends on the mass ratio of the merging  MBHs, we have to model the mass-growth of  
MBHs and the merger efficiency. 
We base our model of MBH growth on the empirical correlation found between MBH masses 
and the properties of their hosts, and on the suggestion that these correlations are 
established during galaxy mergers that fuel MBH accretion and form bulges. 
{We therefore assume that after every merger between two galaxies with a mass ratio
larger than $1:10$, their MBHs climb to the same relation with the velocity
dispersion of the halo, as it is seen today \citep{gultekinetal09b}:
\beq
M=1.3\times10^8 M_\odot \left(\frac{\sigma}{200 \kms} \right)^{4.24}.
\eeq

We link the correlation between the black hole mass  and the central stellar velocity dispersion of the 
host with the empirical  correlation between  the central stellar velocity dispersion and the asymptotic circular 
velocity ($V_{\rm c}$) of galaxies (Ferrarese 2002; see also Pizzella et al. 2005;  Baes et al. 2003). 
\beq
\sigma=200 \kms \left(\frac{V_{\rm c}}{304 \kms}\right)^{1.19}.
\eeq

The latter is a measure of the total mass of the dark matter halo of 
the host galaxies.  A halo of mass $M_{\rm h}$ collapsing at redshift $z$ has a circular velocity
\beq V_{\rm c}= 142 \kms \left[\frac{M_{\rm h}}{10^{12} \ M_{\msun} }\right]^{1/3} 
\left[\frac {\Omm}{\Ommz}\ \frac{\Delta_{\rm c}} {18\pi^2}\right]^{1/6} 
(1+z)^{1/2}  
\eeq 
where $\Delta_{\rm c}$ is the over--density at virialization relative 
to the critical density. 
For a WMAP5 cosmology we adopt here the  fitting
formula (Bryan \& Norman 1998) $\Delta_{\rm c}=18\pi^2+82 d-39 d^2$, 
where $d\equiv \Ommz-1$ is evaluated at the collapse redshift, so
that $ \Ommz={\Omm (1+z)^3}/({\Omm (1+z)^3+\Oml+\Omk (1+z)^2})$. 

Therefore, if a dark matter halo of mass $M_{\rm h}$ at redshift $z$ hosts a MBH, we can
derive the MBH mass via $V_{\rm c}$ and $\sigma$ \citep[see also][]{VHM,Rhook2005, Croton2009}. The relationship 
between black hole and dark matter halo mass is:
\beq
M=1.3\times10^{8} M_\odot \left[\frac{M_h}{9.410^{13} M_\odot} \right]^{5/3}
\left[\frac {\Omm}{\Ommz}\ \frac{\Delta_{\rm c}} {18\pi^2}\right]^{5/6}(1+z)^{5/2}, 
\eeq
in good agreement with the recent result by \cite{Bandara2009}.

We further assume that  MBHs merge within the merger timescale of their host haloes ($t_{\rm merge}$), which is a likely assumption for  MBH binaries formed after gas rich galaxy mergers \citep[and references therein]{dotti07}. We adopt the relations suggested by \cite{Taffoni2003} for the orbital  decay of merging haloes.  We treat as MBH mergers  at a given $z$ those MBH binaries that are expected to merge during that specific timestep. If two haloes start interacting at $z_{\rm in}$, corresponding to a Hubble time $t_H(z_{\rm in})$ with a merger timescale $t_{\rm merge}$,  then we consider these MBHs merged at $z_{\rm fin}$ corresponding to $t_H(z_{\rm fin})=t_H(z_{\rm in})+t_{\rm merge}$. Although we follow the dynamical evolution of 
each MBH along cosmic time, we note that dynamical friction appears to be efficient for mergers with mass ratio of 
the progenitors larger than $1:10$. Small satellites suffer severe mass losses by the tidal
perturbations induced by the gravitational field of the primary halo. This progressive mass loss increases the decay time
that can be of order the Hubble time when the mass ratio is smaller than $1:10$. 

Finally, the spin of MBHs is fixed to be $\hat a=0.9$, in order to obtain upper limits to the recoil consequences.  }

This set of models, though simple, provides a wide range of histories
that can be considered to bracket some extreme behaviours.  
We find that the environment of the MBH population plays an important role. 
Along the merger history of massive galaxies, the fraction of ejected
MBHs decreases more rapidly with decreasing redshift, dropping below
50\% by $z\sim 7$ for $M_0=4\times10^{13}\msun$.  The 50\% threshold is
reached at later times in the `trees' of less massive haloes:
$z\sim5$ for $M_0=2\times10^{12}\msun$ and $z\sim2$ for
$M_0=2\times10^{11}\msun$.  

\begin{figure}
\includegraphics[width=0.45\textwidth,angle=0]{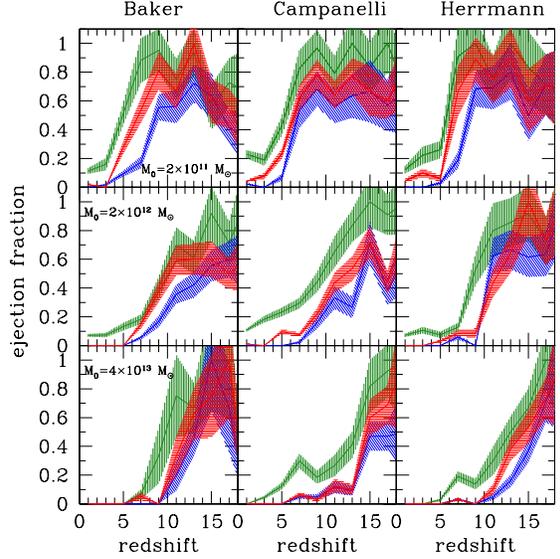}
\caption{Probability of MBH ejections as a function of redshift along
the merger history of the main halo in each merger tree.  Each row
corresponds to different halo masses $M_0$ at $z=0$ as indicated in
the left panels.  Each column as indicated corresponds to the
different fitting function used to calculate recoil speed.  The {\it
green, vertically hatched} regions are for an isotropic distribution
of spin orientations; the {\it red, horizontally hatched} regions are
for the `hot' simulations by Dotti et al.\ (2009); and the {\it blue,
diagonally hatched} regions are for the `cold' simulations by Dotti et
al.\ (2009).  In every panel the MBH spins are fixed at $\hat a=0.9$.
The trends and results are in strong qualitative agreement with the
results from the simpler simulations presented in
Fig.~\ref{fig:ej_peaks}.  There is also general agreement among the
different fitting functions.}
\label{fig:ejectionave_5}
\end{figure}

It is particularly instructive to examine the ejection history of the
tree's main halo, which is the galaxy we would see today
(Figure~\ref{fig:ejectionave_5}).  The main halo is usually among the
most massive haloes in the tree at any time, implying the largest
escape velocities.  The fraction of ejected MBHs is therefore lower.
The central MBH in a large galaxy has a very small probability of
being lost after $z=5$.  Since most of the growth of MBHs happens
between $z=3$ and $z=1$, corresponding to the peak in quasar activity,
we argue that, if accretion is responsible for creating the
correlations between MBHs and their hosts
\citep{sr98,1999MNRAS.308L..39F,2003ApJ...596L..27K}, then MBHs hosted
in large galaxies will sit close to the expected correlation.  The
central MBH in a small galaxy (e.g., $M_0=2\times10^{11}\msun$) has
instead a large probability ($\sim 20$\%) of being ejected all the way
to today.  If such a low-redshift ejection happens, it is not
immediately clear if the galaxy can re-acquire a MBH {\it and} if the
latter can grow to the hypothetical mass that the correlation with the
host would suggest.

The alternative case of MBH formation (haloes with $>10^{10} \msun$)
gives quantitatively similar results for large galaxies, e.g.,
$M_0=4\times10^{13}\msun$, and qualitatively similar results for small
galaxies. Black holes form much later in this model (as the emergence
of more massive galaxies
must wait until
$z \simeq 5-7$), and this pushes the ejection fraction to first rise
steeply, as more and more MBHs are available to merge, and then
decrease as galaxies grow in mass deepening the potential wells.

It is most interesting that the cosmic ejection rate is very similar
for fit B compared to fits CL and H, notwithstanding the different
$\eta$ dependence ($\eta^3$ vs $\eta^2$). The reason is simple. From
Figure~\ref{fig:comparison} it is evident that the mass-ratio
dependence picks up at $q<0.1$ (where, incidentally, recoil velocities
become much smaller than escape velocities from galaxies). However,
mergers with $q\ll1$ are rather uncommon \citep{VHM,Sesanaetal2005},
due to dynamical effects. During a galactic merger, it is dynamical
friction that drags in the satellite, along with its central MBH,
towards the center of the more massive progenitor. When the orbital
decay is efficient, the satellite hole moves towards the center of the
more massive progenitor, leading to the formation of a bound MBH
binary. The efficiency of dynamical friction decreases with the mass
ratio of the merging galaxies: only nearly equal mass galaxy mergers
(`major mergers', mass ratio larger than $\simeq 1:10$) lead to
efficient MBH binary formation within timescales shorter than the
Hubble time \citep{Taffoni2003}.  These effects must be convolved with the mass-ratio
probability distribution. As the mass function of haloes (and galaxies)
is steep, the probability of halo mergers decreases with increasing
mass ratio.  That is, dynamically efficient major mergers are rare,
and minor mergers are common but inefficient at forming MBH binaries.

\begin{figure}
\includegraphics[width=0.45\textwidth,angle=0]{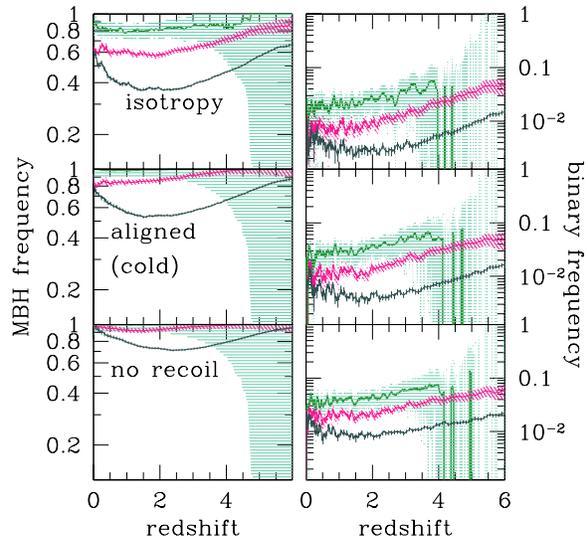}
\caption{ Left: frequency of MBHs in galaxies as a function of
 redshift for haloes above three different mass thresholds: $M>10^{10}
 \msun$ (\emph{gray, vertically hatched}) $M >10^{11} \msun $
 (\emph{magenta, diagonally hatched}) $M>10^{12}\msun$ (\emph{cyan,
 horizontally hatched}) for isotropic spin orientations (\emph{top})
 and spin orientations from the `cold' simulations of Dotti et
 al.~(2009) (\emph{middle}).  The bottom panel shows the black hole
 occupation fraction when the recoil is arbitrarily set to zero at
 each MBH merger. Right: binary MBH frequency for the same models and
 mass thresholds.}
\label{fig:OF}
\end{figure}

We can now derive how strong or mild recoils influence the frequency
of MBHs in galaxies. We analyse here the $\nu_c>3$ case only, as by
construction the alternative case (mass threshold $>10^{10} \msun$)
has an occupation fraction of unity for all haloes above the threshold,
as if a MBH is lost, a new immediately materialises. 
.
We first derive a control case, in which we ignore recoils altogether.
This control case is shown in the bottom panel of
Figure~\ref{fig:OF}. The MBH frequency is calculated above a given
minimum halo mass ($>10^{10}\, \msun$; $>10^{11}\, \msun$; $>10^{12}\,
\msun$).  As discussed by \cite{Menou2001} and \cite{VHM}, the
frequency of MBHs in haloes above a fixed mass threshold initially
decreases with cosmic time as lower mass haloes lacking MBHs become
more massive than the assumed threshold.  Eventually, the occupation
fraction starts to increase as the total number of individual haloes
drops.

Given that different fitting formulae give very similar results, we
show here the results that we obtain using fit CL.  The worst case
scenario, the case with the largest number of ejections, is the
isotropic case, which leads to the largest recoils (see
Figure~\ref{fig:comparison}).  The top panel of Figure~\ref{fig:OF}
indicates that recoils can decrease the overall frequency of MBHs in
small galaxies to $\sim60$\%, while they have little effect on the
frequency of MBHs in large galaxies (at most a 20\% effect).  The
middle panel depicts the cold aligned case. We consider the latter the
more realistic situation for gas-rich mergers, which are expected for
high-redshift galaxies, at least.  It is reassuring to notice that
recoils are not dangerous for most galaxies.

This is the result of a combination of effects: on the one hand MBHs
hosted in small haloes, with shallow potential wells and low escape
velocities, have a high ejection probability
(Figure~\ref{fig:ejectionave_5}), on the other hand the MBH merger
rate is very low along their galaxy formation merger hierarchy: MBH
formation processes are inefficient in such shallow potential wells,
and the anti-hierarchical nature of the galaxy assembly implies that
not much action happens in low-bias systems at high-redshift.  This is
exemplified in the right panels of Figure~\ref{fig:OF}, in which we
show the frequency of MBH close binaries for the same mass thresholds
we used to calculate the MBH frequency (note the different y-axis
scales). {Na\"\i vely}, one would expect the frequency of double MBHs
to scale as the square of the MBH frequency, but the frequency of
close binaries is suppressed with respect to the frequency of pairs
because of the requirement of efficient orbital decay. Our simulations
show that gravitational recoil is not expected to be efficient at
ejecting MBHs with mass $\simeq10^6\msun$.  This mass range is where
the {\it Laser Interferometric Space Antenna} ({\it LISA})
gravitational wave observatory will be most sensitive to extreme mass
ratio inspiral (EMRI) events \citep{gair2009}.  Thus the event rate
for EMRIs is not likely to be affected by recoils.

Summarising, the larger ejection probability that MBHs hosted in
low-bias haloes have is counteracted by their lower merger probability,
thus leading to an overall small change in the MBH frequency as a
function of the recoil strength.

\section{Discussion}
Ejection of MBHs may explain the unusual case of NGC~3115.  NGC~3115
is an S0 galaxy with MBH measured to have mass $M=1\times10^9\msun$
\citep{kr92}, yet the surface-brightness profile of the central bulge
has a small break radius $r_b\sim2\units{pc}$ with a steep inner slope
slope consistent with a power-law $\gamma=0.52$ \citep{laueretal07b}.
This presents a contradiction of theoretical expectations for a galaxy
of this size ($M_{h} \approx 10^{13}\msun$).  

A galaxy as massive as NGC~3115 is expected to have merged frequently
enough to have had a binary black hole at some point in its history
\citep{vmh03}.  According to the standard picture of core scouring
\citep{bbr80, 1997AJ....114.1771F, 2003ApJ...596..860M, laueretal07b},
the binary black hole would eject stars on elongated orbits as orbital
energy is transferred from the binary to kinetic energy in the stars.
This process shrinks the binary separation and flattens the stellar
number density profile so that the surface brightens profile appears
as a core at the center.  The galaxy's core can be described by its
deficit in stellar mass, i.e., the mass in stars ejected from what
was previously a power-law profile.  The most recent numerical
simulations on black hole mergers find that for nearly equal masses,
the mass deficit should scale with the total mass of the binary
\citep{2006ApJ...648..976M}; but for mass ratios far from unity, the
mass deficit should scale with the mass of the secondary
\citep{2008ApJ...686..432S}.  Observationally, the mass deficit is
found to scale linearly with existing $M_\mathrm{BH}$
\citep{laueretal07,2009ApJ...691L.142K}.  Regardless of its magnitude,
the mass deficit is expected to persist after the binary coalesces
because the remnant black hole acts as a `guardian' that prevents
dynamical refilling of the core \citep{laueretal07}.

The observations of NGC~3115, however, reveal that this is not what
happened.  There is an existing MBH at its center that would have
acted as the core's guardian, but the surface brightness profiles show
no core.  There are several possible explanations.  First, core
scouring may be greatly diminished if the final major merger had a
significant amount of gas \citep{Kormendy2009}.  The gas drag speeds
up the process of MBH merger before the MBH binary can eject many
stars.  Even if core scouring occurred, a subsequent gas-rich merger
could re-fill the core by depositing gas that would form stars.  Here,
we suggest another possibility.  The final MBH binary is ejected by
gravitational recoil \citep[or three-body encounters,][]{gmh04,
gmh06,hl07}, but the MBH is replaced in a subsequent merger.  Taking
into consideration the typical mass ratios of merging MBHs as a
function of galaxy bias and cosmic time, the probability that a MBH in
NGC~3115 has been ejected at redshift $z<5$ is 10\% in the case of a
`dry merger', but less than 1\% in the case of a `wet merger'. This is
because at late cosmic times the binary mass ratio distribution
becoming shallow, with $q\ll1$ becoming less probable
\citep{volonteri07}.  In general, in order to have a probability
larger than 10\% that a MBH recoils with a velocity larger than the
escape velocity of the progenitor of NGC~3115 at $z\simeq5$ the mass
ratio of the merging MBHs must be $q>0.3$ for the isotropic and hot
cases (assuming spins $\hat a=0.9$.). For the cold case, even for
$q=1$ and $\hat a=0.9$, the probability is 0.8\%.  In the ejection
scenario NGC~3115 should have had at least one major merger that
involved two MBH-hosting galaxies where substantial star formation or
AGN feedback kept the gas pressurized (corresponding to our hot
simulations).

\section{Summary}
We assessed the effects of spin alignment on the strength of the
gravitational recoil along the cosmic build-up of galaxies and MBHs.
We determined ejection probabilities for mergers in gas-poor galaxies,
where the MBH binary coalescence is driven by stellar dynamical
processes, and the spin-orbit configuration is expected to be
isotropically distributed. We contrast this case of gas-rich mergers,
where we expect MBH spins to align with the orbital angular
momentum. This is because in gas-rich environments MBHs accrete gas,
which exerts gravito-magnetic torques that align the spins of both the
MBHs with the angular momentum of the large-scale gas flow.  We find
that for aligned configurations the ejection probability is strongly
suppressed (by at least a factor of 2). Along the merger history of a
large elliptical, the ejection probability becomes negligible at
$z<5$, while small galaxies have ejection probabilities of order 20\%
even today. However, the MBH merger rate is very low along their
merger hierarchy of small galaxies: MBH formation processes are likely
inefficient in such shallow potential wells.  The occupation fraction
of MBHs, intimately related to halo bias and MBH formation efficiency,
therefore plays a crucial role in increasing the retention fraction.
Recoils can effectively decrease the overall frequency of MBHs in
small galaxies to $\sim60$\%, while they have little effect on the
frequency of MBHs in large galaxies (at most a 20\% effect).

\section*{Acknowledgements}
MV acknowledges support from a Rackham faculty grant.

\appendix
\section{Dependence of the results on the MBH-host relationships}
The estimates presented in the main text assume MBH masses correlate
with the velocity dispersion of their host halo \citep{Bandara2009},
and also that this relationship is redshift--independent.

To test the robustness of our results against the assumed MBH-host
relationship we implement a very different relationship, that is, we
assume that the MBH mass scales with the mass of the dark matter halo:
$M=10^{-6}M_h$. This scaling leads to different dependencies of the
MBH mass ratio, $q$, with the properties of the merging hosts
(redshift, halo mass).  The redshift dependence, however, cancels out
in the expression for the MBH mass ratio, $q$, as long as the MBHs
grow almost coevally. More important is the different scaling of $q$
with the properties of the merging galaxies. In the ``sigma-based"
relationship we have: $q\propto (M_{h,2}/M_{h,1})^{5/3}$, while for
the ``mass-based" relationship $q\propto (M_{h,2}/M_{h,1})$. For
$M_{h,2}/M_{h,1}<1$ the sigma-based scaling leads to lower $q$
values (this is true for any scaling relationship with an exponent
larger than 1), and milder recoils.

We find that the occupation fraction of small galaxies is mostly
unaffected by the choice of the MBH-host relationship, and the
difference for larger galaxies is within 20\%, with the linear mass
scaling leading to a lower occupation fraction. The redshift
dependence of the MBH frequency is almost identical.

\bibliographystyle{mn2e}

\label{lastpage}
\end{document}